\documentclass[a4paper]{article}
\usepackage{amsmath}
\usepackage{amssymb}
\newtheorem{Theorem}{Theorem}
\newtheorem{Proposition}[Theorem]{Proposition}
\newtheorem{Lemma}[Theorem]{Lemma}
\newtheorem{Corollary}[Theorem]{Corollary}
\newtheorem{Definition}[Theorem]{Definition}
\newtheorem{Notation}[Theorem]{Notation}
\def\id{\mathop{\mathrm{id}}}
\def\Id{\mathop{\mathrm{Id}}}
\def\oId{\mathop{\mathrm{Id}^\#}}
\def\Range{\mathop{\mathrm{Range}}}
\def\Tr{\mathop{\mathrm{Tr}}}
\def\MM{{\Lambda}}
\def\dM{{\MM^{\!(\dagger)}}}
\def\fE{{\mathfrak E}}
\def\fM{{\mathfrak M}}
\def\fX{{\mathfrak{X}}}
\def\tM{\tilde{\MM}}
\def\oN{{\otimes N}}
\def\sH{\mathcal{H}}  

\begin{document}
\title{Lower bound for entanglement cost \\of antisymmetric states}
\author{Toshiyuki Shimono\thanks{shimono@is.s.u-tokyo.ac.jp\newline
\indent Department of Computer Scicence,
Graduate School of Information Science and Technology,
University of Tokyo.
}}
\maketitle

\begin{abstract}
This report gives a lower bound of entanglement cost for 
antisymmetric states of bipartite $d$-level systems to be $\log_2 \frac{d}{d-1}$
ebit
(for $d=3, E_c \ge 0.585\ldots $).
The paper \cite{VDC} claims that the value is equal to one ebit 
for $d=3$ , since all of the eigenvalues of reduced matrix of
any pure states affiliating to $\sH_-^{\otimes N}$  is not greater than 
$2^{-N}$
thus the von Neumann entropy is not less than $N$,
 but the proof is not true.
Hence whether the value is equal to or less than one ebit is not clear 
at this moment.
\end{abstract}

\section{Introduction}
Entanglement cost is determined by asymptotic behavior of 
entanglement formation \cite{HHT}, but it is regarded to be very difficult to 
calculate. The paper \cite{VDC} claims that 
the entanglement cost of antisymmetric states of bipartite three-level system 
is 
one ebit. 
However, the proof in that paper is not correct (for the version of January 11, 2002) 
as explained as follows. 
The essential point of the proof of that paper is,
all of the eigenvalues of reduced matrix of
any pure states affiliating to $\sH_-^{\otimes N}$ for $d=3$ is not greater than 
$2^{-N}$.
Thus the von Neumann entropy $- \sum \lambda_i \log_2 \lambda_i$ 
is not less than $N$ bit. 
Hence any mixed states supported on antisymmetric states, whose decomposition is always 
on antisymmetric states, have the entanglement formation not less than $N$ ebit. 
Therefore
entanglement cost is not less than one ebit. 
However, there exist counterexamples, i.e., 
the largest eigenvalue of the reduced matrix of 
$
\frac{1}{\sqrt{12}}
\sum_{i,j=1\dots 3}
\left( |ii\rangle\!_A|jj\rangle_{\!B} -|ij\rangle\!_A|ji\rangle_{\!B} \right)
=
\frac{1}{\sqrt{3}}
\sum_{1\le i < j\le 3} \left(\frac{
|i\rangle\!_A|j\rangle_{\!B} -|j\rangle\!_A|i\rangle_{\!B} 
}{\sqrt{2}}\right)^{\otimes 2}
\in\sH_-^{\otimes 2}
$
is $\frac{1}{3}\left(>\frac{1}{2^2}\right)$.
Hence at this moment, it is not clear 
whether the entanglement cost of antisymmetric states for bipartite
three-level system is one ebit
or not.
\\

\noindent
This report furnish a lower bound of the entanglement cost of 
antisymmteric states for bipartite $d$-level systems.
It is proved that all of the eigenvalues of reduced matrix of any pure
states affiliating to $\sH_-^{\otimes N}$ for general $d$ is not greater than 
$\left(\frac{d-1}{d}\right)^N$.
This is proved by investigating a certain map $\tM$, which is defined at
expression (\ref{note:lambda2}) later, whether it is CP or not. %

%
%
%
\section{Results}
\subsection{Problem Setup}\label{subsection:2.1}
%
Let us assume each of $\sH_A$ and $\sH_B$ is a $d$-dimentional Hilbert space with 
basis $D:=\{|i\rangle\}_{i=1\ldots d}$ and $\sH_{AB}:=\sH_A\otimes\sH_B$.
 For $1\le i \lneqq j \le d$, 
\[ |(i,j)\rangle:=
{\displaystyle \frac{
|i\rangle_A\otimes|j\rangle_B-|j\rangle_A\otimes|i\rangle_B
}{\sqrt{2}} 
}
\quad \in\quad \sH_{AB}
\]
and $D' := \{ |(i,j)\rangle \}_{1\le i \lneqq j \le d } $,
the antisymmetric space $\sH_- := \mathop{\mathrm{span}}D' \subset H_{AB}$ .
\begin{Notation}[matrices]
For a positive integer $m$, $\fM_m$ is a set of
$m\times m$-dimentional matrices with each entry a complex number
$\mathbb{C}$. 
For a set $\fX$,
$\left[a_{ij}\right]_{i,j\in \fX}$
is a matrix with 
$(i,j)$-component specified $a_{ij}$,
and 
$\fM(\fX):=\left\{[a_{ij}]_{i,j\in \fX} | \{a_{ij}\}\subset\mathbb{C}\right\}$
is a collection of matrices each rows and columns are labelled with elements of
$\fX$.
\end{Notation}
\begin{Notation}[partial order between matrices]
The partial order $\le$ in $\fM(\fX)$ is introduced as follows.
For $X_1,X_2\in\fM(\fX)$ , 
$X_1\ge X_2 \Leftrightarrow X_2 \le X_1 \Leftrightarrow X_1-X_2\ge 0
 \Leftrightarrow X_1-X_2$ 
is a positive matrix. Note that a positive matrix is always hermitian.
\end{Notation}
\begin{Definition}[$\MM:\fM(D')\rightarrow\fM(D)$]
The map $\MM:\fM(D')\rightarrow\fM(D)$ is defined as follows. 
First $X\in\fM(D') $, is regarded as a antisymmetric state
$\rho_1 :=\sum_{I,J \in D} X_{IJ}|I\rangle\langle J| \in \sH_-$, Then 
$\rho_1$ is reduced into $\sH_A$ by the operation
$\rho_2 := \Tr\limits_B \rho_1\in \sH_A$, and is converted into the matrix 
representation $Y\in\fM(D)$ with basis $D$ 
satisfying $\rho_2=\sum_{i,j\in D}Y_{ij}|i\rangle \langle j|$.
This transformation $X\mapsto Y$ is the map $\MM$.
\end{Definition}
%
The derivations of this map $\MM$ are investigated in the section \ref{subsection:2.2}.
%
%
%
%
\begin{Notation}[$\fE_{ij}^\fX$]
For a set $\fX$ and $i,j\in \fX$,
$\fE_{ij}^\fX\in\fM(\fX)$
is a matrix with entry $1$ only at $(i,j)$-component and $0$ 
elsewhere. For example,
for $\fX=\{1,2,3\}$,
 $\fE_{1,2}^\fX = \left(\begin{smallmatrix} 0&1&0\\0&0&0\\0&0&0
 \end{smallmatrix}\right)$.
\end{Notation}
The notations such as 
$
[\fE^\fX_{IJ}]_{I,J\in \fX}
$, which is equal to
\[
\begin{pmatrix} 
\fE^\fX_{11}&\fE^\fX_{12}&\fE^\fX_{13}\\\fE^\fX_{21}&\fE^\fX_{22}&\fE^\fX_{23}\\\fE^\fX_{31}&\fE^\fX_{32}&\fE^\fX_{33}
 \end{pmatrix}
=
\left(\begin{smallmatrix} 
1&0&0&&0&1&0&&0&0&1\\0&0&0&&0&0&0&&0&0&0\\0&0&0&&0&0&0&&0&0&0\\\\
0&0&0&&0&0&0&&0&0&0\\1&0&0&&0&1&0&&0&0&1\\0&0&0&&0&0&0&&0&0&0\\\\
0&0&0&&0&0&0&&0&0&0\\0&0&0&&0&0&0&&0&0&0\\1&0&0&&0&1&0&&0&0&1
 \end{smallmatrix}\right) 
\]
when $\fX$ is $\{1,2,3\}$ , for example, will be used in this report. This example indicates a $3\times 3$ block matrix with $3\times 3$ matrices, and can be treated as a $9\times 9$. 

%
%
\begin{Notation}[$\dM$]
$\dM$ is difined as a mapping
$X\mapsto \MM(X^\dagger)$ 
that is a compound map $\MM$ preceeded by matrix adjoint(Hermitan transpose).
Note that $\dM$ operates on hermitian matrices as same as $\MM$ operate on,
i.e. for a hermitian matrix $X$, $\dM(X)=\MM(X)$ since $X^\dagger=X$.
\end{Notation}

\noindent In this report "map" is a mapping between matrices.
\begin{Notation}[identities]
\noindent 
Let us assume each of $\fM,\fM'$ is either of $\fM_m$ or $\fM(\fX)$.
Then $\id\limits_\fM,\Id\limits_\fM,\oId\limits_{\fM,\fM'}$ are denoted as follows:
$\id\limits_\fM$  is an identity matrix of $\fM$, 
$\Id\limits_\fM$  is an identity map on $\fM$, 
$\oId\limits_{\fM,\fM'} $  is a linear map $\fM \ni X \mapsto (\Tr X) \cdot\id\limits_\fM \in \fM' $.
$\fM,\fM'$ will be dropped sometimes,
such as, $\id,\Id$ and $\oId$.
\end{Notation}
%
%
\subsection{Propositions and theorems}\label{subsection:2.2}
%
%
%
\begin{Lemma}\label{lem:1}
For scalars $x,y$, eigenvalues of 
$ \displaystyle 
  \left(
    \Id\limits_{\fM(D')}   \otimes \Big( x\MM + y\dM \Big)
  \right)
 \left[\fE^{D'}_{IJ}\right]_{I,J\in D'}
$
are $ -y,\frac{1}{2}y,\frac{d-1}{2}x+\frac{1}{2}y $ .

%
\end{Lemma}
{\bf proof}
The considering matrix is equal to 
$\Xi := 
 \left[
\Big( x\MM + y\dM \Big)
\fE^{D'}_{IJ}
\right]_{I,J\in D'}
$
.
 For $(i,j),(k,l)\in D'$,
$
\MM \left( \fE^{D'}_{(i,j)(k,l)} \right)
= \Tr\limits_B |(i,j)\rangle\langle(k,l)|
= \frac{1}{2} 
\begin{smallmatrix}
     &
{{}_k \qquad {}_l} 
    \\
  \begin{smallmatrix} {\phantom |}^i \\{\phantom |}^j\end{smallmatrix}\!\!
    &
  \left(\begin{smallmatrix}
    \delta_{jl}&-\delta_{jk}\\-\delta_{il}&\delta_{ik}
  \end{smallmatrix}\right)
    \\
 &   \phantom{    {{}_i \qquad {}_l}  }
\end{smallmatrix}
$
where $\delta$ is Kronecker's delta, and
$
\dM \left( \fE^{D'}_{(i,j)(k,l)} \right)
= \frac{1}{2} 
\begin{smallmatrix}
     &
{{}_i \qquad {}_j} 
    \\
  \begin{smallmatrix} {\phantom |}^k \\{\phantom |}^l\end{smallmatrix}\!\!
    &
  \left(\begin{smallmatrix}
    \delta_{jl}&-\delta_{il}\\-\delta_{jk}&\delta_{ik}
  \end{smallmatrix}\right)
    \\
 &   \phantom{    {{}_i \qquad {}_l}  }
\end{smallmatrix}
$
.
Observing the whole matrix $\Xi$, it is decomposed 
into the form of direct product 
$\displaystyle 
\Xi = \frac{y}{2}\Xi_1 \oplus \left(\frac{x}{2}\Xi_2+\frac{y}{2}\Xi_3\right)$ where \\ 
\begin{eqnarray*}
\Xi_1 
&=& \mathop{\bigoplus}\limits^{i,j,k}_{1\le i < j < k \le d	}
\begin{smallmatrix}
     &
{{}_{(i,j)\otimes k} \, {}_{(i,k)\otimes j}\, {}_{(j,k)\otimes i}} 
    \\
  \begin{smallmatrix}
 {\phantom |}^{(i,j)\otimes k} \\
 {\phantom |}^{(i,k)\otimes j} \\
 {\phantom |}^{(j,k)\otimes i}
\end{smallmatrix}\!\!
    &
  \left(\begin{matrix}    0&1&-1\\1&0&1\\-1&1&0   \end{matrix}\right)
    \\
 &   \phantom{    {{}_i \qquad {}_l}  }
\end{smallmatrix}
, \\
\Xi_2 
&=& \mathop{\bigoplus}\limits^{i}_{1\le i  \le d	}
\bordermatrix{
                               &       &        &        &        &        &        \cr
{\phantom x}^{(1,i)\otimes i}  &1      & \cdots & 1      & -1     & \cdots & -1     \cr
\quad          \vdots          &\vdots &        & \vdots & \vdots &        & \vdots \cr
{\phantom x}^{(i-1,i)\otimes i}&1      & \cdots & 1      & -1     & \cdots & -1     \cr
{\phantom x}^{(i,i+1)\otimes i}&-1     & \cdots & -1     & 1      & \cdots & 1      \cr
   \quad        \vdots         &\vdots &        & \vdots & \vdots &        & \vdots \cr
{\phantom x}^{(i,d)\otimes i}  &-1     & \cdots & -1     & 1      & \cdots & 1   
}
, \\
\Xi_3 
&=& \mathop{\bigoplus}\limits^{i}_{1\le i  \le d	}
\bordermatrix{
                               &       &        &        &        &        &        \cr
{\phantom x}^{(1,i)\otimes i}  &1      &        &        &        &        &        \cr
\quad          \vdots          &       & \ddots &        &        &   0    &        \cr
{\phantom x}^{(i-1,i)\otimes i}&       &        & 1      &        &        &        \cr
{\phantom x}^{(i,i+1)\otimes i}&       &        &        & 1      &        &        \cr
   \quad        \vdots         &       &   0    &        &        & \ddots &        \cr
{\phantom x}^{(i,d)\otimes i}  &       &        &        &        &        & 1   
}
.
\end{eqnarray*}
$ \frac{y}{2}\Xi_1 $ has eigenvalues $-y,\frac{y}{2}$ , and 
$ \left(\frac{x}{2}\Xi_2+\frac{y}{2}\Xi_3\right)$  has eigenvalues 
$\frac{1}{2}y , \frac{d-1}{2}x+\frac{1}{2}y$ .

\begin{flushright}$\blacksquare$\end{flushright}  
\begin{Lemma}
Let $\lambda({x,y}) := 
\max \{ |-y|,|\frac{1}{2}y|,|\frac{d-1}{2}x+\frac{1}{2}y| \}$  .
Then 
\begin{eqnarray*}
\quad\mathop{\mathrm{arg\,min}}\limits^{(x,y)}_{x+y=1} \lambda({x,y}) &=& 
\left(\frac{1}{d},\frac{d-1}{d}\right)
, \\
\min\limits_{x+y=1} \lambda({x,y}) &=& 
\lambda\left(\frac{1}{d},\frac{d-1}{d}\right) = \frac{d-1}{d} 
.
\end{eqnarray*} 
\end{Lemma}
\begin{Notation}[$\tilde\lambda,\tM$]
Let 
\begin{eqnarray}
\tilde\lambda &:=&\frac{d-1}{d} \label{note:lambda1}
,
\\
\tM &:=& \frac{1}{d} \MM + \frac{d-1}{d} \dM \label{note:lambda2}
.
\end{eqnarray} 
\end{Notation}
Note that due to the last two lemmata,
\begin{equation}\label{ineq:lambda}
  - \tilde\lambda \id 
                    \le 
  \left(
    \Id\limits_{\fM(D')}   \otimes \tM
  \right)
 \left[\fE^{D'}_{IJ}\right]_{I,J\in D'}
                    \le 
    \tilde\lambda \id 
\end{equation}
i.e. the absolute values of every eigenvalues of the central side are not larger than 
(\ref{note:lambda1}). Note that (\ref{note:lambda2}) operates on hermitian matrices as same as $\MM$ operates on.
\begin{Notation}
If it is written ${D'}^N$ where $D'$ is the set of basis of $\sH_-$, 
it associates the direct product of the set ${D'}$,
indicates the basis of $\sH_-^\oN$.
\end{Notation}
%
%
%
\begin{Lemma}\label{lem:2}
\begin{equation}\label{eq:2} 
\left( \Id\limits_{\fM(D')^\oN} \otimes 
\left(\tilde\lambda^N \oId - \tM^\oN  \right) \right)
\left[\fE^{{D'}^N}_{IJ}\right]_{I,J\in {D'}^N}
\ge 0
\end{equation}
This is equivalent to that the left hand side is positive matrix.
Here, $\oId$ is a map from 
$ \fM(D')^\oN $ into $ \fM(D)^\oN $.
\end{Lemma}
{\bf proof} \phantom{a}
 The inequality (\ref{eq:2}) is equivalent to
\[
\Big( \Id\otimes\tilde\lambda^N \oId \Big)
\left[\fE^{{D'}^N}_{IJ}\right]
\ge 
\Big( \Id\otimes\tM^\oN  \Big)
\left[\fE^{{D'}^N}_{IJ}\right]
\]
To verify this inequality, the following is enough.%
\[
\left\{
\begin{array}{ll}
\mathit{LHS.} &=
\tilde\lambda^N \left[\oId 
\big(\fE^{{D'}^N}_{IJ}\big)\right]_{I,J}
=\tilde\lambda^N\left[\delta_{IJ}\id\right]_{I,J} 
= \tilde\lambda^N \id
\\
\mathit{RHS.}&=
\Big( \mathop{\Id^\oN}\limits_{\fM(D')}  \otimes \tM^\oN   \Big)
\Big(\big[\fE^{{D'}}_{IJ}\big]{}^\oN\Big)
=
\Big( \big(\Id\limits_{\fM(D')} \otimes \tM \big)
\big[\fE^{{D'}}_{IJ}\big]\Big)^\oN
\\
&\le
\left(\tilde\lambda \id \right)^\oN = 
\tilde\lambda^N \id
\qquad ( \le \mbox{  {\it due to }} (\ref{ineq:lambda}  ) \quad)
\end{array}
\right.
\]
\begin{flushright}$\blacksquare$\end{flushright}

\noindent The last lemma successively induces next two propositions.
\begin{Proposition}
$\; \displaystyle 
\tilde\lambda^N
\oId - \tM^\oN   $  is a CP map.
\end{Proposition}
This is due to (\ref{eq:2}) and \cite{C}. 
Namely, 
for a map $\Gamma:\fM_m\to\fM_n$ , it is CP iff
$\left(\Id_{\fM_m}\otimes\Gamma\right)\left[\fE_{ij}\right]_{i,j=1\ldots m}
=\left[\Gamma\left(\fE_{ij}\right)\right]_{i,j=1\ldots m}$ is a positive matrix.
\begin{Proposition}\label{prop:positivity}
$\displaystyle \; \MM^\oN(X) \le 
\tilde\lambda^N\id \;$ 
 for $X \in \fM(D')^{\oN}$ if
$ X \ge 0 \;, \;\Tr X=1$.
\end{Proposition}

\noindent
Let we denote entanglement measures $E,E_f$ and $E_c$ as von Neumann entropy,
entanglement formation and entanglement cost, respectively.
Proposition \ref{prop:positivity} is applied to calculate (a lower bound of)
these values.
In this report, the base of entropy is always fixed to two, regardless the
dimension $d$ of either $\sH_A$ or $\sH_B$. That is to say, for density matrix
$\rho$, the von Neumann entropy is $E(\rho) = -\Tr \rho \log_2 \rho$,
not $-\Tr \rho \log_d \rho$.
\begin{Theorem}\label{Th:1}
$E(|\Psi\rangle)\ge N \log_2\frac{d}{d-1}$
 for any pure state $|\Psi\rangle \in \sH_-^\oN$.
\end{Theorem}
This is because the last proposition indicates that all of the eigenvalues
of the reduced matrix from arbitary antisymmetric states are less than or
equal to $\left({d/d-1}\right)^{-N}$.

\begin{Lemma}
$E_f(\sigma)\ge N\log_2{\frac{d}{d-1}}$ 
for any density matrix $\sigma$ supported on 
$\sH_-^\oN$.
\end{Lemma}
{\bf proof }
\noindent Entanglement formation is defined as
\begin{equation}\label{eq:ef}
 \displaystyle E_f(\rho):=\min_{\bigl(p_i,|\Phi_i\rangle\bigr)_i \in \Delta(\rho)}\sum_i p_i E(\Phi_i) 
\end{equation}
where 
\begin{equation}
 \Delta(\rho):=\left\{\bigl(p_i,|\Phi_i\rangle\bigr)_i \Big|
(p_i>0, \|\Phi_i\|=1)\forall i,
\sum_i p_i = 1, 
\sum_i p_i |\Phi_i\rangle\langle \Phi_i|=\rho\right\}
\end{equation}
is the collection of all possible decompositions of $\rho$.
It is known that all of $|\Phi_i\rangle$ 
induced from $\Delta(\rho)$ satisfy 
$|\Phi_i\rangle\in\Range(\rho)$ , where
$\Range(\rho)$ is sometimes called 
image space of a matrix $\rho$ 
which is a collection of $\rho|\psi\rangle$ with $|\psi\rangle$ 
running over the domain of $\rho$.
Hence 
\begin{equation}
E_f(\rho)\ge \min \{ E(\Phi) |
 \Phi \in \Range(\rho),\|\Phi\|=1\}
.
\end{equation}
The condition of the lemma above implies $\mathop\mathrm{Range}(\rho)\subseteq H_-^\oN$ , 
therefore the last theorem implies $E_f(\sigma)\ge N$.
\begin{flushright}$\blacksquare$\end{flushright}
\hspace{0pt}\\
\noindent 
The paper \cite{HHT} claims that 
$\displaystyle E_c(\rho)=\lim_{N\to\infty}\frac{E_f(\rho^{\otimes N})}{N}$ ,
therefore the value of entanglement cost is given as follows.
\begin{Theorem}
$E_c(\sigma)\ge N
\log_2\frac{d}{d-1}
$ for any density matrix $\sigma$ supported on 
$\sH_-^\oN$.
\end{Theorem}
\begin{Corollary}[The lower bound of entanglement cost for $\sH_-$]
 \label{Cor:1}
\begin{equation}\label{eq:main}
E_c(\sigma)\ge \log_2\frac{d}{d-1}
\end{equation} 
for any density matrix $\sigma$ supported on 
$\sH_-$.
\end{Corollary}
\section{Conclusion and Discussion}
This report gave a lower bound of entanglement cost of antisymmetric states 
for $d$-dimentional antisymmetric states as inequality (\ref{eq:main}). However, 
it is still open probem whether the entanglement cost for $d=3$ is one ebit or not is not clear.\\

%
%
%
%
\noindent
\\
\noindent{\bf Acknowledgements}
\\
\noindent I am greatful to Dr.~W.~Y.~Hwang, Dr.~H.~Fan for introducing to me 
the paper \cite{VDC} and discussions, Dr.~K.~Matsumoto and Dr.~M.~Hamada for giving me
discussions and references, and Prof.~H.~Imai for supporting me 
the occasions.

\end{document}